\documentclass[journal=nalefd,manuscript=letter,layout=twocolumn]{achemso}

\usepackage[version=3]{mhchem} 
\usepackage{graphicx}
\usepackage{dcolumn}
\usepackage{bm}
\usepackage[mathlines]{lineno}
\usepackage{amssymb}
\usepackage{amsmath}
\usepackage{natbib}
\usepackage{soul,color}

\title{Two-dimensional
       phosphorus carbide:\\
       Competition between $sp^2$ and $sp^3$ bonding}

\author{Jie Guan}
\affiliation{Physics and Astronomy Department,
             Michigan State University,
             East Lansing, Michigan 48824, USA}
\author{Dan Liu}
\affiliation{Physics and Astronomy Department,
             Michigan State University,
             East Lansing, Michigan 48824, USA}
\author{Zhen Zhu}
\affiliation{Physics and Astronomy Department,
             Michigan State University,
             East Lansing, Michigan 48824, USA}
\altaffiliation{Materials Department,
             University of California,
             Santa Barbara, California 93106, USA}
\author{David Tom\'{a}nek}
\affiliation{Physics and Astronomy Department,
             Michigan State University,
             East Lansing, Michigan 48824, USA}
\email{tomanek@pa.msu.edu}

\abbreviations{2D, TMD, DFT, PBE, DOS}


\begin{document}


\begin{abstract}
We propose previously unknown allotropes of phosphorus carbide
(PC) in the stable shape of an atomically thin layer. Different
stable geometries, which result from the competition between
$sp^2$ bonding found in graphitic C and $sp^3$ bonding found in
black P, may be mapped onto 2D tiling patterns that simplify
categorizing of the structures. Depending on the category, we
identify 2D-PC structures that can be metallic, semi-metallic with
an anisotropic Dirac cone, or direct-gap semiconductors with their
gap tunable by in-layer strain.
\end{abstract}



\begin{figure*}[t]
\includegraphics[width=1.6\columnwidth]{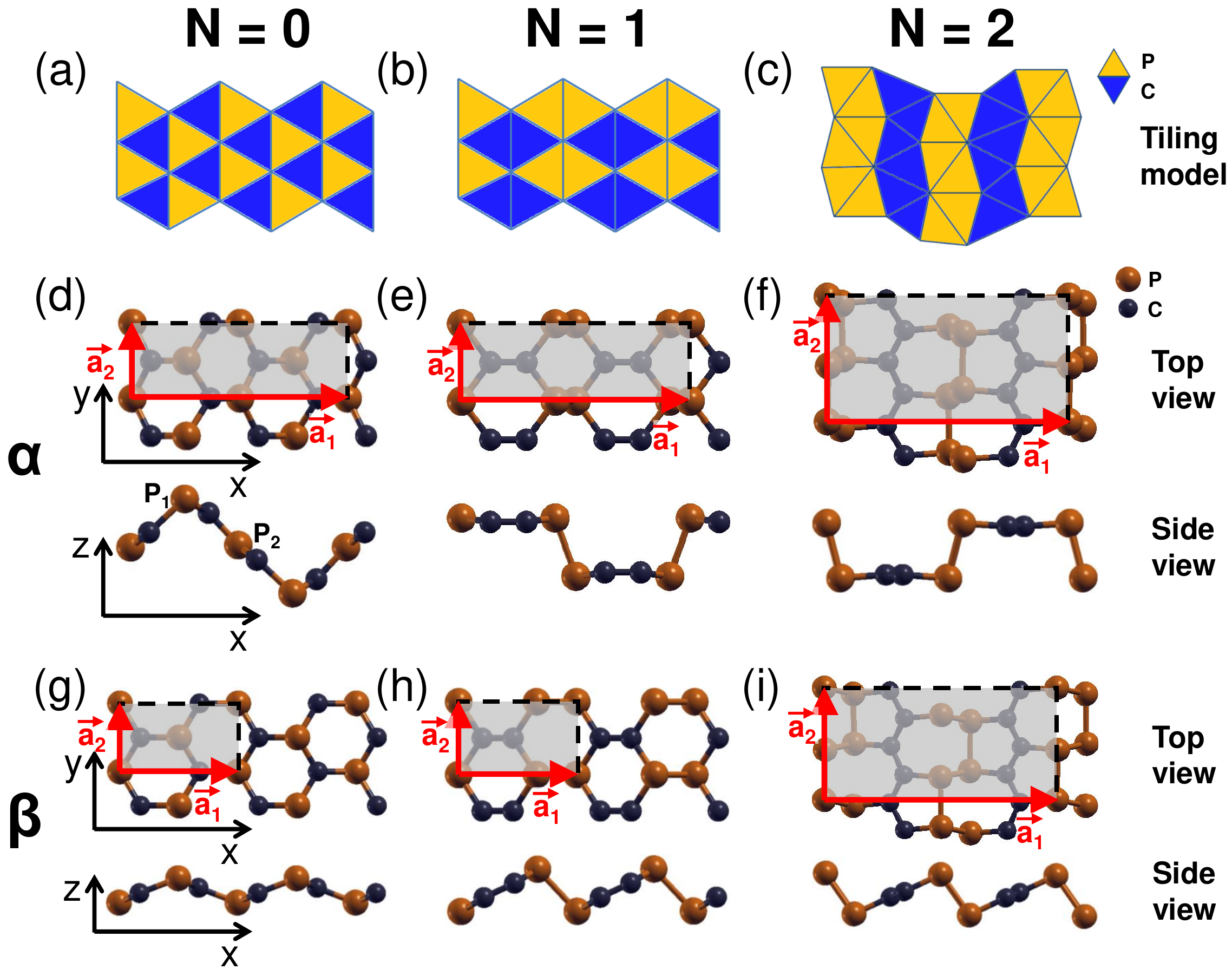}
\caption{(Color online) Possible stable structures of an
atomically thin PC monolayer, represented by (a-c) a tiling
pattern and (d-i) by ball-and-stick models in both top and side
view. The number of like nearest neighbors defines the structural
category $N$. There are two stable allotropes, $\alpha$ and
$\beta$, for each $N$. The primitive unit cells are highlighted
and the lattice vectors are shown by red arrows. Two inequivalent
P sites are distinguished by a subscript in (d). \label{fig1}}
\end{figure*}

There is growing interest in 2D semiconductors, both fundamental
and as potential components in flexible, low-power electronic
circuitry. A large number of substances with unique advantages and
limitations has been studied in this respect, but consensus has
not been reached regarding the optimum candidate. Semi-metallic
graphene with an excellent carrier mobility has received the most
attention so far, but all attempts to open up a sizeable, robust,
and reproducible band gap have failed due to the negative side
effects of the different modifications%
~\cite{{graphane},{PKimPRL07},{GNR-Fujita96},{DT190}}. Transition
metal dichalchogenides (TMDs) such as MoS$_2$~\cite{{Kis2011},{Fuhrer2013}}
or TcS$_2${~\cite{TcS2EK}}
do have a sizeable
fundamental band gap, but a lower carrier
mobility. Recently isolated
few-layer films of black phosphorus, including phosphorene
monolayers, combine high carrier mobility with a sizeable and
tunable fundamental band gap%
~\cite{{DT229},{Li2014}}, but have limited stability in
air~\cite{Hersam-phosphorene14}.

Since both elemental carbon and phosphorus form stable 2D
monolayers, which have been studied extensively, it is intriguing
to find out, whether the compound phosphorus carbide (PC), %
also called carbon phosphide,
may also be stable as a monolayer and display properties that may
even be superior to both constituents. The plausibility of a 2D
structure of PC derives from the same three-fold coordination
found both in graphene and phosphorene. On the other hand, the 2D
structure will likely suffer from a competition between the planar
$sp^2$ bonding characteristic of graphene and the significantly
different non-planar $sp^3$ bonding found in phosphorene. The
postulated 2D structure of PC with 1:1 stoichiometry is
fundamentally different from the amorphous structure observed in
deposited thin solid films~\cite{Furlan2013}, the postulated
foam-like 3D structure~\cite{Hart2010}, or the postulated
GaSe-like multi-layer
structures of PC containing C and P with the same $sp^3$ hybridization%
~\cite{{Claeyssens2004},{Zheng2003prb}}.
On the other hand, 2D allotropes of PC are somehow related to
postulated and partly to observed fullerene-like structures of %
CP$_x$~{\cite{{FL-CP1},{FL-CP2}}} %
and CN$_x$
{~\cite{{GueorguievCPL05a},{GueorguievCPL05b},%
{SjostromPRL95},{NeidhardtJAP03b}}}, %
and to g-C$_3$N$_4$, called graphitic carbon
nitride.{~\cite{g-c3n4}}%


In this Letter, based on {\em ab initio} density functional
calculations, we propose previously unknown allotropes of
phosphorus carbide in the stable shape of an atomically thin
layer. We find that different stable geometries, which result from
the competition between $sp^2$ bonding found in graphitic C and
$sp^3$ bonding found in black P, may be mapped onto 2D tiling
patterns that simplify categorizing of the structures. We
introduce the structural category $N$, defined by the number of
like nearest neighbors, and find that $N$ correlates with the
stability and the electronic structure characteristic. Depending
on the category, we identify 2D-PC structures that can be
metallic, semi-metallic with an anisotropic Dirac cone, or
direct-gap semiconductors with their gap tunable by in-layer
strain.

\section{Results and Discussion}


As mentioned above, all atoms in the 2D-PC allotropes are
threefold coordinated, similar to the planar honeycomb lattice of
graphene. Thus, the structure can be topologically mapped onto a
2D lattice with sites occupied either by P or C atoms. Bisecting
all nearest-neighbor bonds by lines yields a 2D tiling pattern,
where each triangular tile with a characteristic color represents
either a P or a C atom. Next, we define a structural category $N$
for each allotrope, with $N$ given by the number of like nearest
neighbors. For $N=0$, none of the atoms are connected to any like
neighbors. Each C or P atom has only one like (C or P) neighbor
for $N=1$, and two like neighbors for $N=2$. There is no $N=3$
structure, which would imply a pure carbon or phosphorus lattice.
The tiling patterns for different 2D-PC allotropes are shown in
Fig.~\ref{fig1}(a)-\ref{fig1}(c). A similar categorization scheme
has been used previously to distinguish between different
allotropes of 2D phosphorene~\cite{DT240}, where $N$ was the
number of ``like'' neighbors either in the upper or lower position
within the lattice.

\begin{figure}[t]
\includegraphics[width=1.0\columnwidth]{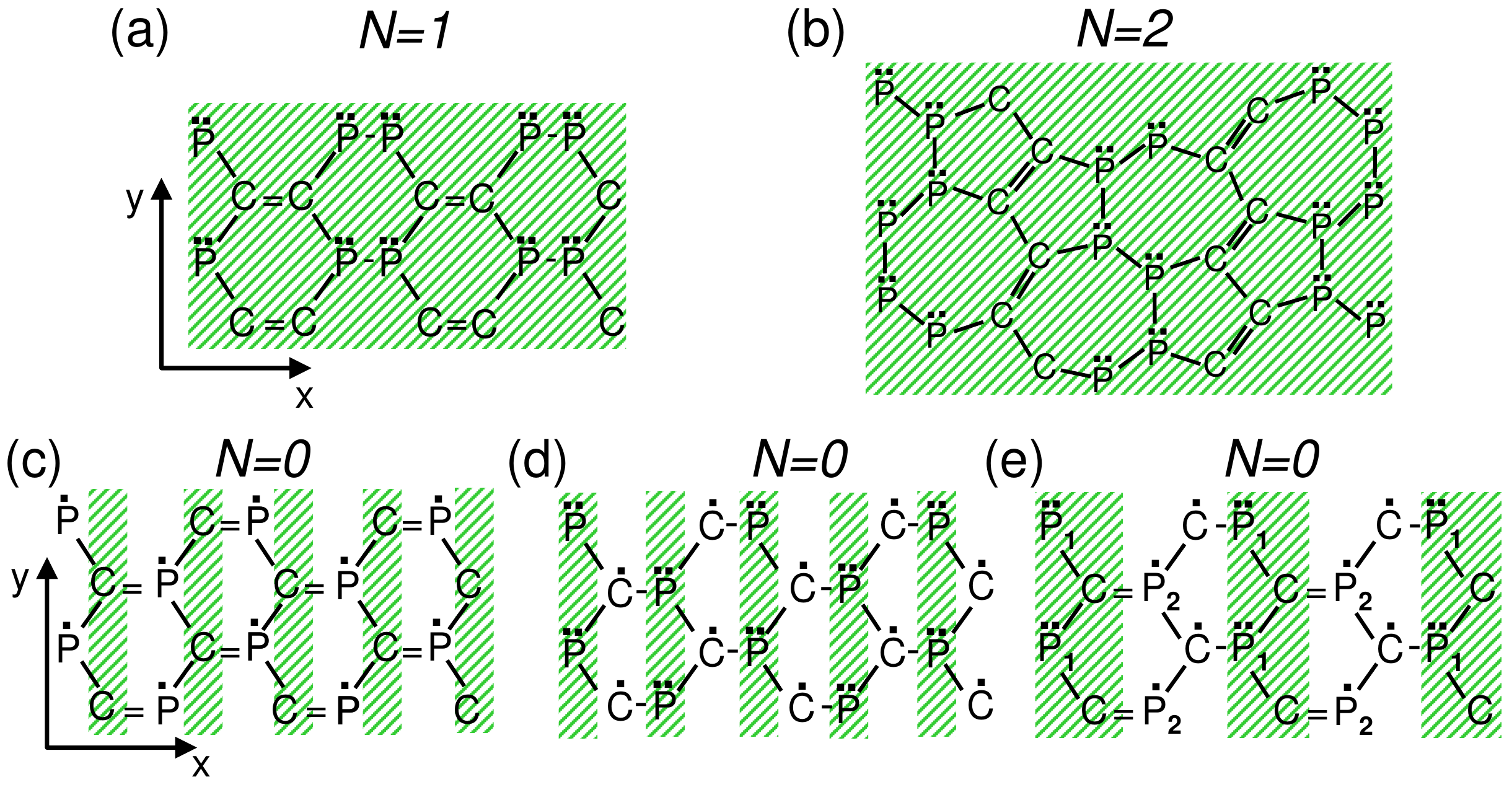}
\caption{(Color online) Bonding configuration in (a) $N=1$, (b)
$N=2$, and (c-e) $N=0$ category 2D-PC allotropes. Green-shaded
regions indicate sites that satisfy the octet rule discussed in
the text. Bonding in $\beta_0$-PC is characterized by panel (d)
and bonding in $\alpha_0$-PC is described by panel (e).
\label{fig2}}
\end{figure}

Whereas the tiling pattern is useful for simple categorization, it
does not provide information about the nontrivial optimum
structure shown in Fig.~\ref{fig1}(d)-\ref{fig1}(i), which results
from a competition between the favored planar $sp^2$ hybridization
of C and non-planar $sp^3$ hybridization of P. The side view of
structures displayed in Fig.~\ref{fig1} best illustrates that
allotropes with the same value of $N$ may be structurally
different. In analogy to the different postulated phosphorene
allotropes~\cite{{DT230},{DT232}}, we distinguish $\alpha_N$,
which display a black-P-like armchair structure in side view, from
$\beta_N$ phases of PC, which display a blue-P-like (or
grey-As-like) zigzag structure in side view, and use the index $N$
to identify the structural category.

We start our discussion with $N=1$ allotropes $\alpha_1$-PC and
$\beta_1$-PC, shown in the middle column in Fig.~\ref{fig1}.
According to the definition of $N$, each atom has one neighbor of
the same species and two of different species, forming isolated
P-P and C-C dimers, as seen in the tiling pattern and the atomic
structures. As seen in Fig.~\ref{fig2}(a), the chemical octet
rule~\cite{octet-rule} is satisfied both on C sites in %
the graphitic %
$sp^2$ configuration and on P sites, containing a lone electron
pair, in $sp^3$ configuration, indicating stability. Both
allotropes have rectangular unit cells consisting of distorted
hexagons. The unit cell of $\alpha_1$-PC with 8 atoms is larger
than that of $\beta_1$-PC with four atoms.

In $N=2$ allotropes $\alpha_2$-PC and $\beta_2$-PC, shown in the
right column of Fig.~\ref{fig1}, each atom has two like neighbors
and one unlike neighbor. In the side view, these allotropes look
very similar to those of the $N=1$ category. The main difference
becomes apparent in the top view. %
Whereas $N=1$ structures contain ethylene-like C$_2$ units that
are interconnected by P$_2$ dimers, $N=2$ systems contain
contiguous trans-polyacetylene-like all-carbon chains that are
separated by P-chains. %
Due to the difference between the locally planar $sp^2$ bonding of
C atoms and locally non-planar $sp^3$ bonding of P atoms, and due
to the difference between equilibrium C-C and P-P bond lengths,
the hexagons found in $N=1$ structures change to pentagon-heptagon
pairs in the optimum
$N=2$ %
structure
resembling pentheptite or haeckelite structures related to
graphitic carbon.
As seen in Fig.~\ref{fig2}(b), similar to $N=1$ structures, the
chemical octet rule is satisfied on both C and P sites. The
lattice of $\alpha_2$-PC and $\beta_2$-PC allotropes contains
rectangular unit cells with sixteen atoms. %

\begin{table*}[t]
\caption{Calculated properties of different 2D-PC allotropes.
$<E_{coh}>$ is the cohesive energy per ``average'' atom with
respect to isolated atoms. $<{\Delta}E>=<E_{coh}>-<E_{coh,max}>$
describes the relative stability of a system with respect to the
most stable structure. $|\vec{a_1}|$ and $|\vec{a_2}|$ are the
in-plane lattice constants defined in Fig.~\protect\ref{fig1}.
$d_{P-P}$, $d_{P-C}$ and $d_{C-C}$ are the equilibrium bond
lengths between the respective species. In $\alpha_0$-PC, the
P$_1$-C bonds differ from the P$_2$-C bonds in length. }
\begin{tabular}{|l|c|c|c|c|c|c|}
\hline %
Structure                  & $\alpha_0$-PC & $\beta_0$-PC %
                           & $\alpha_1$-PC & $\beta_1$-PC %
                           & $\alpha_2$-PC & $\beta_2$-PC \\%
\hline %
$<E_{coh}>$~(eV/atom)      & 4.80          & 4.75 %
                           & 5.05          & 5.06 %
                           & 5.20          & 5.20 \\ %
$<{\Delta}E>$~(eV/atom)    & $-0.40$       & $-0.45$ %
                           & $-0.15$       & $-0.14$ %
                           & 0.00          & 0.00 \\%
$|\vec{a}_1|$~({\AA})      & 8.41          & 5.12 %
                           & 8.73          & 4.76 %
                           & 9.84          & 10.59 \\ %
$|\vec{a}_2|$~({\AA})      & 2.94          & 2.95 %
                           & 2.95          & 2.95 %
                           & 5.11          & 5.11 \\ %
$d_{P-P}$~({\AA})          & --            & -- %
                           & 2.36          & 2.36 %
                           & 2.29          & 2.29 \\%
$d_{P-C}$~({\AA})          & 1.86~(P$_1$)  & 1.78 %
                           & 1.84          & 1.84 %
                           & 1.85          & 1.85 \\%
                           & 1.71~(P$_2$)  &  %
                           &               &  %
                           &               &  \\%
$d_{C-C}$~({\AA})          & --            & -- %
                           & 1.38          & 1.38 %
                           & 1.44          & 1.44 \\%
\hline
\end{tabular}
\label{table1}
\end{table*}

In 2D PC compounds of category $N=0$, shown in the left column of
Fig.~\ref{fig1}, each atom is surrounded by three unlike
neighbors. There is no bonding configuration that would satisfy
the octet rules on all sites. The bonding configuration depicted
in Fig.~\ref{fig2}(c) satisfies the octet rule only at the C
sites, whereas the configuration in Fig.~\ref{fig2}(d) favors only
the P sites. The bonding configuration depicted in
Fig.~\ref{fig2}(e) contains alternating P-C chains containing P
sites with lone electron pairs and C atoms in $sp^2$
configuration, which satisfy the octet rule, and P-C chains that
do not satisfy it. In whatever bonding arrangement, the bonding
configuration in $N=0$ structures is frustrated. As a consequence,
the $\alpha_0$-PC structure converts spontaneously from an initial
armchair configuration, similar to $\alpha_1$-PC and
$\alpha_2$-PC, to the zigzag structure depicted in
Fig.~\ref{fig1}(d), with details about the structural
transformation discussed in the Supporting Information. The final
$\alpha_0$-PC structure with inequivalent P$_1$ and P$_2$ sites
reflects the bonding configuration in Fig.~\ref{fig2}(e)
containing P$_1$ sites with lone electron pairs and P$_2$ sites
with lone electrons. The $\beta_0$-PC structure, depicted in
Fig.~\ref{fig1}(g), remains locally stable in the electronic
configuration shown in Fig.~\ref{fig2}(d).

Structural characteristics and the binding energy of the different
allotropes are summarized in Table~\ref{table1}. Our energy
results are obtained using the DFT-PBE functional (including spin
polarization where required), which is known to overbind to some
degree. We define the cohesive energy per atom, $<E_{coh}>$, by
dividing the total atomization energy by the total number of
atoms, irrespective of species. The energy values in the first
rows indicate that for given $N$, the $\alpha$ and $\beta$ phases
are almost equally stable, confirming that categorizing structures
by the number of like neighbors at any site makes sense in terms
of stability. Clearly, $N=2$ systems are most stable, followed by
$N=1$ and $N=0$ allotropes. In particular, the cohesive energy of
$N=2$ monolayers exceed the $5.14$~eV/atom value of the postulated
GaSe-like PC multi-layer structures%
~\cite{{Claeyssens2004},{Zheng2003prb}} by $0.06$~eV/atom.

The lower stability of $N=0$ systems has been anticipated above,
since the octet rule can not be satisfied at all sites. We also
note that the $\alpha_0$ phase is slightly more stable than the
$\beta_0$ phase of PC. The stability advantage of $\alpha_0$-PC
derives from the larger variational freedom within the unit cell,
which allows to distinguish two different P sites (P$_1$ and
P$_2$), as shown in Fig.~\ref{fig1}(d) and Fig.~\ref{fig2}(e). The
$\alpha_0$-PC structure consists of P$_1$($sp^3$)-C($sp^2$)
chains, which obey the octet rule and form stable ridges,
alternating with P$_2$-C chains, which do not obey the octet rule
and form terraces.

Additional support for the plausibility of the bonding
configuration depicted in Fig.~\ref{fig2} comes from the
equilibrium bond lengths, which are listed in Table~\ref{table1}.
With the exception of $N=0$ structures, the bond lengths depend
primarily on $N$ and are rather insensitive to the phase ($\alpha$
or $\beta$). For $N=1$ and $N=2$ structures, the C-C bond lengths
lie close to the $1.42$~{\AA} value in $sp^2$ bonded graphite (or
graphene) and the P-P bond lengths are close to the
$2.26-2.29$~{\AA} range found in layered black phosphorus (or
phosphorene).

\begin{figure*}[t]
\includegraphics[width=1.6\columnwidth]{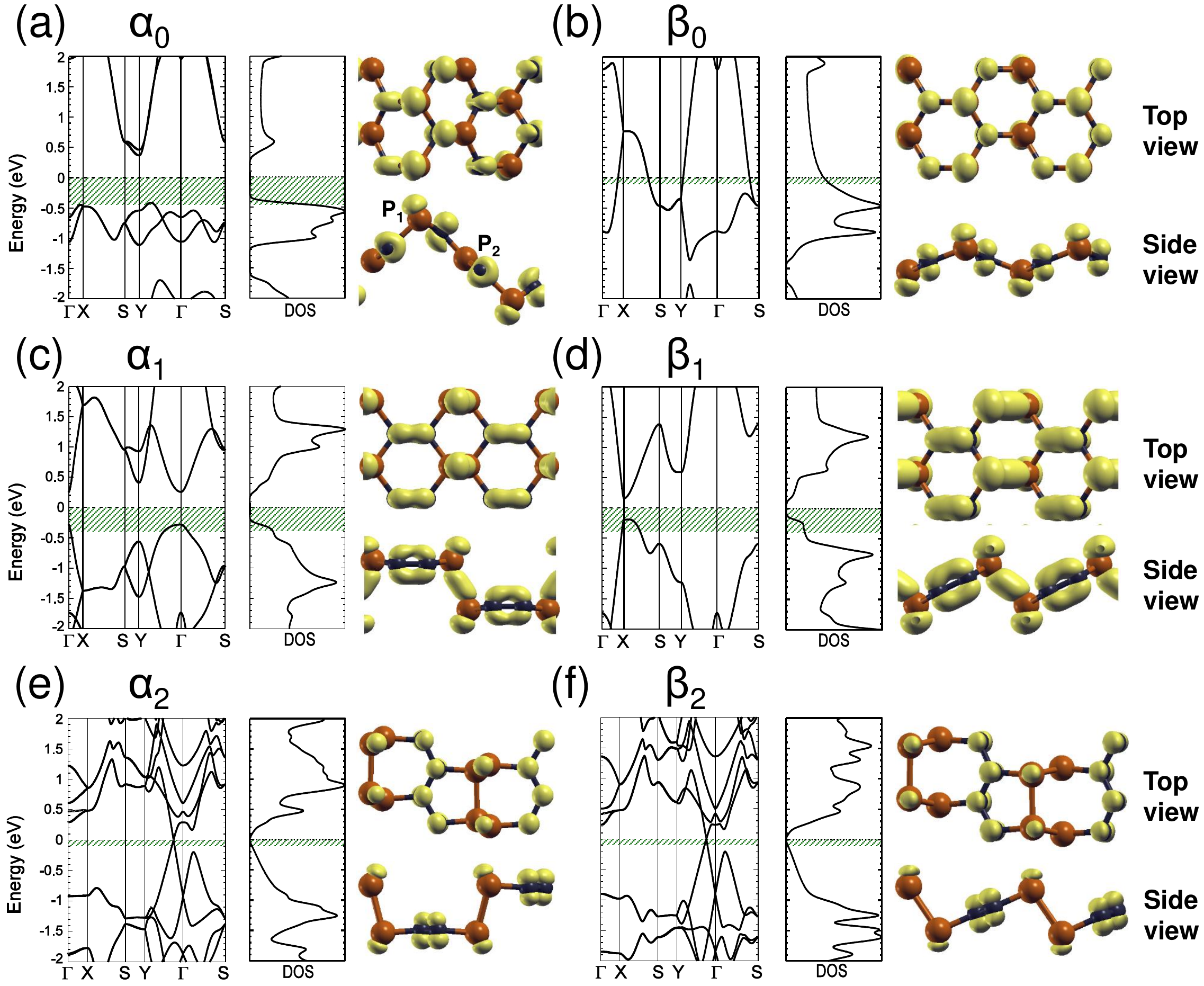}
\caption{(Color online) Electronic band structure, density of
states (DOS), and charge density $\rho_{vc}$ associated with
valence frontier states of $\alpha_N$ and $\beta_N$ allotropes,
where $N$ is the structural category defined in the text and used
in Fig.~\protect\ref{fig1}. The energy range associated with
$\rho_{vc}$ is indicated by the green shaded region in the band
structure and DOS panels and extends from %
$E_F-0.45$~eV$<E<E_F$ for semiconducting $\alpha_0$-PC in (a), %
                                        from %
$E_F-0.40$~eV$<E<E_F$ for semiconducting $\alpha_1$-PC in (c)  %
                                        and $\beta_1$-PC  in (d), %
                                        and from
$E_F-0.10$~eV$<E<E_F$ for metallic $\beta_0$-PC  in (b), and for
                                                 semi-metallic %
                                   $\alpha_2$-PC in (e) and %
                                   $\beta_2$-PC  in (f). %
For each system, isosurface plots of $\rho_{vc}$ are displayed in
the right-side panels and superposed with ball-and-stick models of
the structure in top and side view. The isosurface values of
$\rho_{vc}$ are
$1.0{\times}10^{-3}$~e/{\AA}$^3$ in (a), %
$2.0{\times}10^{-3}$~e/{\AA}$^3$ in (b), %
$0.5{\times}10^{-3}$~e/{\AA}$^3$ in (c) and (d), and %
$0.5{\times}10^{-4}$~e/{\AA}$^3$ in (e) and (f). \label{fig3}}
\end{figure*}

As seen in Fig.~\ref{fig2}(a) and \ref{fig2}(b), P and C atoms are
connected by a single-bond with $d_{\rm P-C}{\approx}1.85$~{\AA}
in $N=1$ and $N=2$ category structures. As suggested above, the
bonding is frustrated at least in parts of $N=0$ structures. In
the significantly reconstructed $\alpha_0$-PC system, depicted in
Fig.~\ref{fig1}(d), we can distinguish P$_1$ sites at ridges from
P$_2$ sites at terraces. The lengths of the three P-C bonds are
very similar at each of the these P sites, but differ
significantly between P$_1$ and P$_2$. At P$_1$ sites that satisfy
the octet rule, as seen in Fig.~\ref{fig2}(e), the P$_1$-C bond
length of $1.86$~{\AA} is very similar to $N=1$ and $N=2$
structures. At P$_2$ sites, which do not satisfy the octet rule,
the frustrated bonds are much shorter with $d_{P-C}=1.71$~{\AA}.
As seen in Fig.~\ref{fig1}(g), there is no reconstruction in the
$\beta_0$-PC structure. As seen in the corresponding
Fig.~\ref{fig2}(c) or \ref{fig2}(d), the octet rule is only
satisfied at either the P or the C sites. The P-C bonds are
frustrated and their length of $1.78$~{\AA} lies in-between the
P$_1$-C and P$_2$-C bond lengths in $\alpha_0$-PC.

\begin{figure}[t]
\includegraphics[width=0.80\columnwidth]{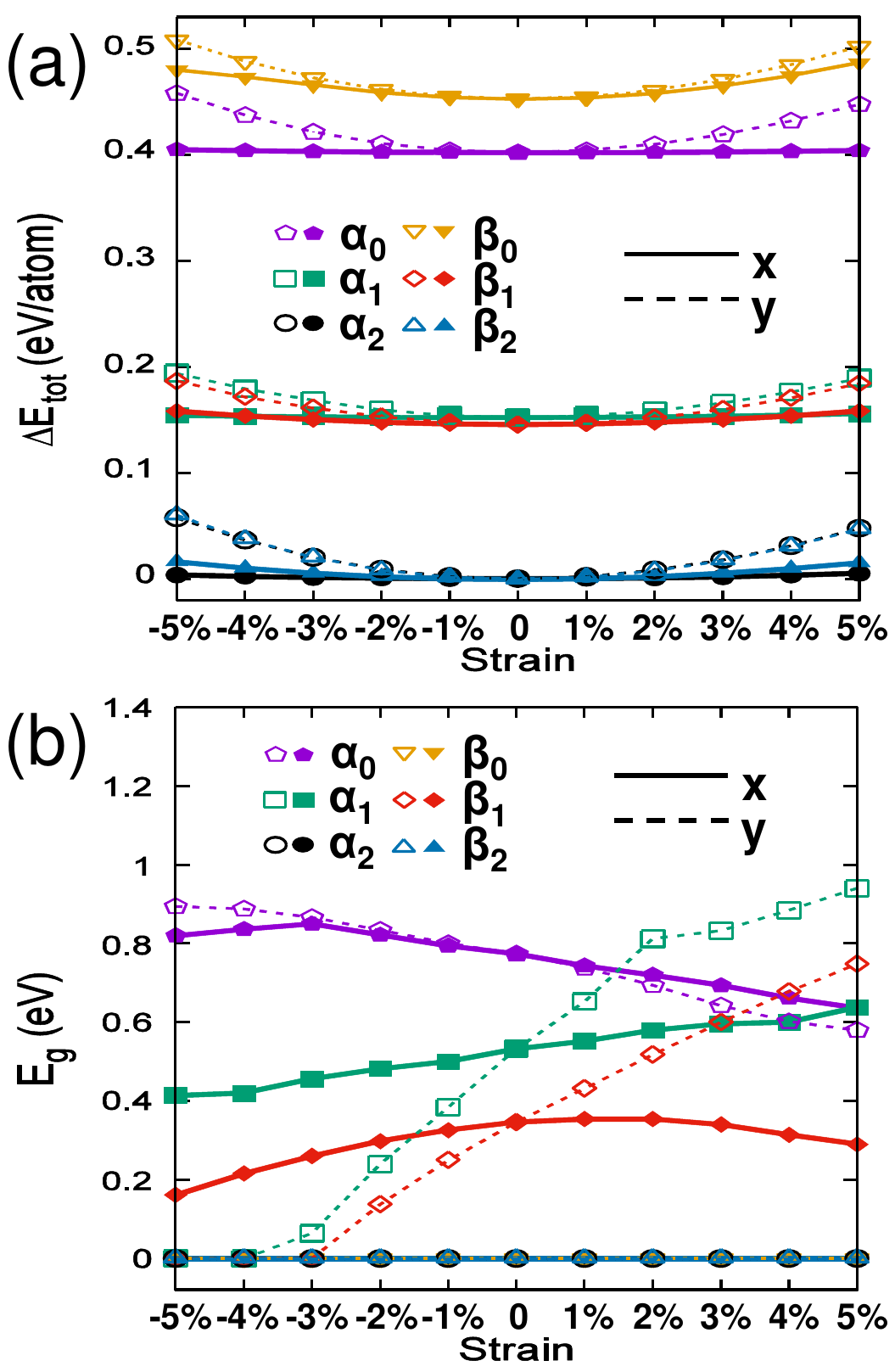}
\caption{(Color online) Effect of uniaxial in-layer strain on (a)
the relative binding energy ${\Delta}E_{tot}$ and (b) the
fundamental band gap in different PC allotropes. Results for
$\alpha_0$-PC, $\beta_0$-PC, $\alpha_1$-PC, $\beta_1$-PC,
$\alpha_2$-PC and $\beta_2$-PC are distinguished by color and
symbols. Results for strain in the $x$-direction, defined in
Fig.~\protect\ref{fig1}, are shown by solid lines and for strain
in the $y$-direction by dashed lines. \label{fig4}}
\end{figure}


Results of our DFT-PBE electronic band structure calculations for
monolayers of the six proposed PC allotropes are presented in
Fig.~\ref{fig3}.

The electronic band structure and associated density of states
(DOS) of $N=0$ systems is shown in Fig.~\ref{fig3}(a) and
\ref{fig3}(b). Our results in Fig.~\ref{fig3}(a) suggest that
$\alpha_0$-PC is an indirect-gap semiconductor with a band gap of
${\approx}0.7$~eV. In stark contrast, the structurally similar
$\beta_0$-PC allotrope is metallic according to
Fig.~\ref{fig3}(b). As suggested earlier, all bonds and electronic
configurations are frustrated in $\beta_0$-PC, with all C sites
engaging only three valence electrons in $sp^2$-like bonds,
leaving one lone electron behind, and the angle at the P ridge
being too large for typical $sp^3$ bonding. This finding, in
particular the presence of a non-bonding electron in the
C$2p_\perp$ orbital, is seen in the frontier states of
$\beta_0$-PC that are depicted in the right panel of
Fig.~\ref{fig2}(b).

$\alpha_0$-PC is quite different from $\beta_0$-PC, as it contains
two inequivalent P and C sites. The P$_1$ site at the ridge
displays the favored $sp^3$ bonding characteristic and its lone
pair orbital is present in the frontier state displayed in the
right-hand panel of Fig.~\ref{fig3}(a). In contrast, the bonding
is very different at the P$_2$ site, where the lone pair orbital
does not contribute to the frontier state. The flat bonding
geometry near this site is reminiscent of $sp^2$ bonding at the C
sites. The added flexibility provided by a larger unit cell allows
for additional stabilization of $\alpha_0$-PC due to the opening
of a band gap, with vague analogy to the Peierls instability.

As seen in Fig.~\ref{fig3}(c) and \ref{fig3}(d), both
$\alpha_1$-PC and $\beta_1$-PC have a direct band gap, %
which we attribute to the presence of isolated ethylene-like units
mentioned above.
The two allotropes display a very similar charge distribution in
their valence frontier states, which contain the lone pair
orbitals on P sites and reflects $sp^2$ bonding between C sites.
The main difference between the two structures is that the 0.4~eV
wide gap in $\alpha_1$-PC is at the $\Gamma$ point, whereas the
0.3~eV gap in $\beta_1$-PC is at the $X$ point. In both
structures, the band dispersion is rather anisotropic near the top
of the valence and bottom of the conduction band, which causes an
anisotropy in the effective mass. We find the effective mass of
both electrons and holes to be much smaller along $x$-direction
than along the $y$-direction, which is
reminiscent of the situation in black phosphorene%
~\cite{{DT229},{Li2014}}. The effective mass anisotropy offers a
significant advantage in transport, since it combines high
mobility of carriers with a large DOS near the band edges.

According to Fig.~\ref{fig3}(e) and \ref{fig3}(f), also the two
$N=2$ allotropes, $\alpha_2$-PC and $\beta_2$-PC, share very
similar band structure, DOS and frontier orbitals due to
structural similarities. The electronic structure of these systems
is nevertheless very different from the other two categories, %
chiefly due to the presence of trans-polyacetylene-like chains
mentioned above.
Both $\alpha_2$-PC and $\beta_2$-PC display a Dirac cone at the
Fermi level, at a crystal momentum between $\Gamma$ and $Y$. As
mentioned before, the distinguishing feature of $N=2$ structures
is the alternation between chains consisting of pure P or pure C
atoms. Fig.~\ref{fig2}(b) indicates that all P sites have occupied
lone pair orbitals, which are also reflected in the frontier
states. The P chains form ridges within the structure, with bond
angles characteristic of $sp^3$ bonding found in black phosphorus.
The structure of the carbon chains, also illustrated in
Fig.~\ref{fig2}(b), resembles that of conjugated
trans-polyacetylene or graphene with $sp^2$ bonding, and the
presence of C$2p_\perp$ orbitals in the frontier states is clearly
seen in the right-side panels of Fig.~\ref{fig3}(e) and
\ref{fig3}(f). Differences between equilibrium bond length and
bond angles of the P and C chains are accommodated by introducing
pentagon-heptagon pairs. The conjugation within C chains and their
suppressed dimerization caused by their bonding to adjacent P
chains lies behind the formation of the Dirac cone. Due to the
strong anisotropy in the system, caused by the direction of the %
trans-polyacetylene-like chains, the Dirac cone is anisotropic in
the plane of the layer. We have found that uniaxial strain may be
used to eliminate the anisotropy of the Dirac cone, but will also
change the location of the Dirac point along the ${\Gamma}-Y$
line. %
More information about the Dirac cone is provided in the
Supporting Information.

Even though DFT-PBE calculations notoriously underestimate the
fundamental band gap between occupied and unoccupied states, the
calculated dispersion $E(${\bf k}$)$ of individual bands is
believed to closely resemble experimental
values. For the sake of comparison, we have also performed DFT-HSE06%
~\cite{{HSE03},{HSE06}} calculations with a hybrid
exchange-correlation functional for the same structures. As seen
in the Supporting Information, our DFT-PBE and DFT-HSE06 results
are closely related. In particular, DFT-HSE06 opens the band gap
in semiconducting $\alpha_0$-PC, $\alpha_1$-PC and $\beta_1$-PC
structures, but keeps the metallic character of $\beta_0$-PC and
the semi-metallic character of $\alpha_2$-PC and $\beta_2$-PC.

Similar to other non-planar 2D systems like phosphorene, PC is
susceptible to even minute in-plane stress, which can cause major
distortions in the geometry, affecting the electronic structure
and bonding. To quantify this effect, we have determined the
effect of tensile and compressive strain on the stability and the
fundamental band gap in the different PC allotropes and present
the results in Fig.~\ref{fig4}. We have considered uniaxial strain
along the $x$- and the $y$-direction, defined in Fig.~\ref{fig1}. %
Since all allotropes discussed here are non-planar, applying
in-layer strain changes the effective thickness of the layers and
vice versa. As expected, layer thickness is reduced under tensile
strain and increased under compressive strain. For strain values
below 5\%, we have observed changes in layer thickness of up to
10\%.
The distinct structural anisotropy, best seen in the side views,
translates into a distinct anisotropy of the strain energy with
respect to the strain direction, shown in Fig.~\ref{fig4}(a).
Similar to black phosphorene, the system appears soft when
strained along the $x$-direction normal to the ridges and valleys,
whereas it is much stiffer when distorted along the $y$-direction.
We find the $\alpha$ phase to be particularly soft in the
$x$-direction, with compressive or tensile strain requiring
${\Delta}E{\lesssim}$5~meV/atom in strain energy.

The dependence of the fundamental band gap on the in-layer strain,
as obtained by our DFT-PBE calculations, is shown in
Fig.~\ref{fig4}(b). We find that compression along the soft
$x$-direction does not affect the band gap much, quite unlike what
is expected to occur in black phosphorene~\cite{DT229}. This is
quite different from our results for strain along the stiffer
$y$-direction. There, we observe the fundamental band gap to
disappear at compressive strain exceeding 4\% for $\alpha_1$-PC
and 3\% for $\beta_1$-PC. We also find that the metallic character
of $\beta_0$-PC and semi-metallic character of $\alpha_2$-PC and
$\beta_2$-PC are not affected by tensile or compressive strains up
to 5\% applied along the $x$- or the $y$-direction. %
Since vertical strain causing a 10\% reduction of the layer
thickness is equivalent to an effective tensile in-layer strain
below 5\%, we can judge its effect on the electronic structure
based on the above findings. %


Even though the cohesive energy of the 2D structures presented
here exceeds that of previously discussed PC systems, the
calculated cohesive energy of 
per formula unit still falls $0.54$~eV short of the sum of the
cohesive energies of pure black phosphorene,
$3.27$~eV, and pure graphene,
$7.67$~eV. Even though the PC allotropes discussed here are all
stable, as seen in the vibration spectra presented in the
Supporting Information, the slight energetic preference for pure
components in favor of the PC compound should offer challenges in
the synthesis. We believe that recent advances in supramolecular
assembly may solve this problem. Similar to our requirements,
precisely designed structures including
graphdiyne~\cite{{graphdiyne1},{graphdiyne2}}, graphene
nanoribbons~\cite{gnr2010} and carbon nanotubes~\cite{Fasel2014}
have been assembled using wet chemical processes from specific
molecular precursors. In the same way, we expect that the
postulated 2D-PC structures may be formed of proper molecular
precursors that contain $sp^2$ bonded carbon and $sp^3$ bonded
phosphorus.

\section{Conclusions}

In conclusion, we have performed {\em ab initio} density
functional calculations and identified previously unknown
allotropes of phosphorus carbide (PC) in the stable shape of an
atomically thin layer. We found that different non-planar stable
geometries, which result from the competition between $sp^2$
bonding found in graphitic C and $sp^3$ bonding found in black P,
may be mapped onto 2D tiling patterns that simplify categorizing
of the structures. We have introduced the structural category $N$,
defined by the number of like nearest neighbors ranging from 0 to
2, and found that $N$ correlates with the stability and the
electronic structure characteristic. We found structures of the
$N=0$ category to be either metallic, or to reconstruct
spontaneously to a more stable structure with a larger unit cell
and a sizeable fundamental gap. Systems of the $N=1$ category are
more stable than $N=0$ systems, display a significant, direct band
gap and a significant anisotropy of the effective mass of
carriers. Category $N=2$ systems are the most stable of all, are
semi-metallic, and display an anisotropic Dirac cone at the Fermi
level. Due to their non-planar character, all systems can sustain
in-layer strain at little energy cost. The fundamental band gap is
not very sensitive to strain in most systems with the exception of
$N=1$ allotropes, where it closes upon applying compressive strain
of ${\lesssim}5$\% along the ridges and valleys.

\section{Methods}

We use {\em ab initio} density functional theory (DFT) as
implemented in the {\textsc SIESTA}~\cite{SIESTA} code to obtain
insight into the equilibrium structure, stability and electronic
properties of 2D-PC allotropes reported in the main manuscript.
Periodic boundary conditions are used throughout the study, with
monolayers represented by a periodic array of slabs separated by a
15~{\AA} thick vacuum region. We use the Perdew-Burke-Ernzerhof
(PBE)~\cite{PBE} exchange-correlation functional, norm-conserving
Troullier-Martins pseudopotentials~\cite{Troullier91}, and a
double-$\zeta$ basis including polarization orbitals. The
reciprocal space is sampled by a fine grid~\cite{Monkhorst-Pack76}
of $8{\times}12{\times}1$~$k$-points in the Brillouin zone of the
primitive unit cell of 4 atoms or its equivalent in supercells.
{\textsc SIESTA} calculations use a mesh cutoff energy of $180$~Ry
to determine the self-consistent charge density, which provides us
with a precision in total energy of ${\leq}2$~meV/atom. All
geometries have been optimized using the conjugate gradient
method~\cite{CGmethod}, until none of the residual
Hellmann-Feynman forces exceeded $10^{-2}$~eV/{\AA}. Since the
fundamental band gap is usually underestimated in DFT-PBE
calculations, we have resorted to the {\textsc HSE06}%
~\cite{{HSE03},{HSE06}} hybrid exchange-correlation functional, as
implemented in the {\textsc VASP}~\cite{VASP,VASP2,VASP3,VASPPAW}
code, to get a different (possibly superior) description of the
band structure. We use $500$~eV as energy cutoff and the default
mixing parameter value $\alpha=0.25$ in these studies. DFT-PBE and
DFT-HSE06 band structure results are compared in the Supporting
Information.

\begin{suppinfo}
The geometry transformation pathway and corresponding changes in
the electronic band structure of $\alpha_0$-PC, details of the
electronic band structure in all PC allotropes, and the phonon
band structure of $\beta_0$-PC and $\beta_1$-PC.\\
\end{suppinfo}
\quad\par

{\noindent\bf Author Information}\\

{\noindent\bf Corresponding Author}\\
$^*$E-mail: {\tt tomanek@pa.msu.edu}

{\noindent\bf Notes}\\
The authors declare no competing financial interest.

\begin{acknowledgement}
We thank Teng Yang and Baojuan Dong for their help in performing
the HSE calculations and acknowledge useful discussions with
Garrett B. King. This study was supported by the NSF/AFOSR EFRI
2-DARE grant number \#EFMA-1433459. Computational resources have
been provided by the Michigan State University High Performance
Computing Center.
\end{acknowledgement}

%

\providecommand{\latin}[1]{#1}
\providecommand*\mcitethebibliography{\thebibliography} \csname
@ifundefined\endcsname{endmcitethebibliography}
  {\let\endmcitethebibliography\endthebibliography}{}

\end{document}